# Triple-junction solar cells with 39.5% terrestrial and 34.2% space efficiency enabled by thick quantum well superlattices

Ryan M. France,[1,2,*] John F. Geisz,[1] Tao Song,[1] Waldo Olavarria,[1] Michelle Young,[1] Alan Kibbler,[1] and Myles A. Steiner[1]

[1]National Renewable Energy Laboratory, Golden CO 80401
[2]Lead Contact
*Correspondence: ryan.france@nrel.gov

## ABSTRACT

Multijunction solar cell design is guided by both the theoretical optimal bandgap combination as well as the realistic limitations to materials with these bandgaps. For instance, triple-junction III-V multijunction solar cells commonly use GaAs as a middle cell because of its near-perfect material quality, despite its bandgap being higher than optimal for the global spectrum. Here, we modify the GaAs bandgap using thick GaInAs/GaAsP strain-balanced quantum well (QW) solar cells with excellent voltage and absorption. These high-performance QWs are incorporated into a triple-junction inverted metamorphic multijunction device consisting of a GaInP top cell, GaInAs/GaAsP QW middle cell, and lattice-mismatched GaInAs bottom cell, each of which has been highly optimized. We demonstrate triple-junction efficiencies of 39.5% and 34.2% under the global and space spectra, respectively, which are higher than previous record six-junction devices.

## INTRODUCTION

High efficiency solar cells are a critical tool in the transition toward renewable energy and the mitigation of climate change and drive many associated applications. III-V tandem solar cells, in particular, exhibit the highest efficiencies of any materials system and are well-suited to area-constrained terrestrial applications such as aerial vehicles or automobiles;[1,2] photoelectrochemical water splitting;[3] generation of hydrogen and other solar fuels;[4] thermophotovoltaic applications such as energy storage;[5] laser power converters;[6] and are leveraging for concentrator photovoltaic systems where light is focused onto the semiconductor to both further increase efficiency and reduce the required semiconductor area.[7,8] III-V tandems are also the dominant technology for satellites and space vehicles,[9-12] where high launch costs make efficiency and weight the key metrics and the radiation environment demands a tolerance to high energy particles.[13-18]

One-sun (non-concentrator) III-V multijunction efficiency has climbed in recent years through improvements to material quality and by adding junctions to reduce thermalization losses while targeting an optimal bandgap combination.[19-21] Improvements to lattice-matched material quality led to record single junction GaAs solar cells,[22,23] high performance GaInP[24] and record GaInP/GaAs tandem solar cells,[25] and a new understanding of photon recycling and luminescent coupling.[26,27] Additional junctions have been added either through advanced metamorphic epitaxy,[28-32] enabling lattice-mismatched alloys with low dislocation density over a wide bandgap range, or wafer bonding,[33,34] where materials grown on multiple host substrates are bonded together. These techniques have led to the recent general progression of record efficiencies of one-sun III-V multijunction solar cells with three- to six-junctions, shown in Fig. 1A, converting up to 39.2% of the global spectrum.[35,36] Although these broad-spectrum devices can be highly efficient, devices with many junctions add complexity to the structure, involve challenging materials, and can be very sensitive to changes in the spectrum. Notably, efficiency gains by adding junctions are progressively smaller in devices with more than three junctions due to practical materials challenges and their intent for concentrator systems.

Here, we demonstrate triple-junction devices with higher efficiency than previous record six-junction devices by improving the bandgap combination of the three-junction inverted metamorphic (IMM) design, thus breaking the paradigm of increasing efficiency by adding junctions. In the original triple-junction IMM design,[37] the top cell was lattice-matched $Ga_{.51}In_{.49}P$ (1.8-1.9 eV), the middle cell was lattice-matched GaAs (1.4 eV), and the bottom cell was high-quality lattice-mismatched GaInAs (1.0 eV). Although lattice-mismatched material has dislocations, metamorphic epitaxy confines most dislocations to inactive regions via facile glide within a compositionally graded buffer, thus resulting in minimal non-radiative recombination in photoactive regions.[29,38] Recent versions of this cell have demonstrated efficiencies of 37.9% under the global spectrum.[39-41] Despite its excellent efficiency, GaAs does not have the ideal bandgap for this design. The ideal bandgap combination of three-junction cells is modeled in Fig. 1 for both the global and AM0 space spectra. The top two bandgaps are plotted in Fig. 1; the bottom cell bandgap is optimized for each combination of the top two cells, as in an IMM, shown in Fig. S6. A material with a lower bandgap than GaAs would be better for both global and space spectra, providing a route towards efficiency improvement. This ideal three-junction bandgap combination was targeted previously using two metamorphic junctions,[42] but the voltage loss in metamorphic 1.35 eV GaInAs resulted in a greater loss in efficiency than was gained by the improved bandgap combination.

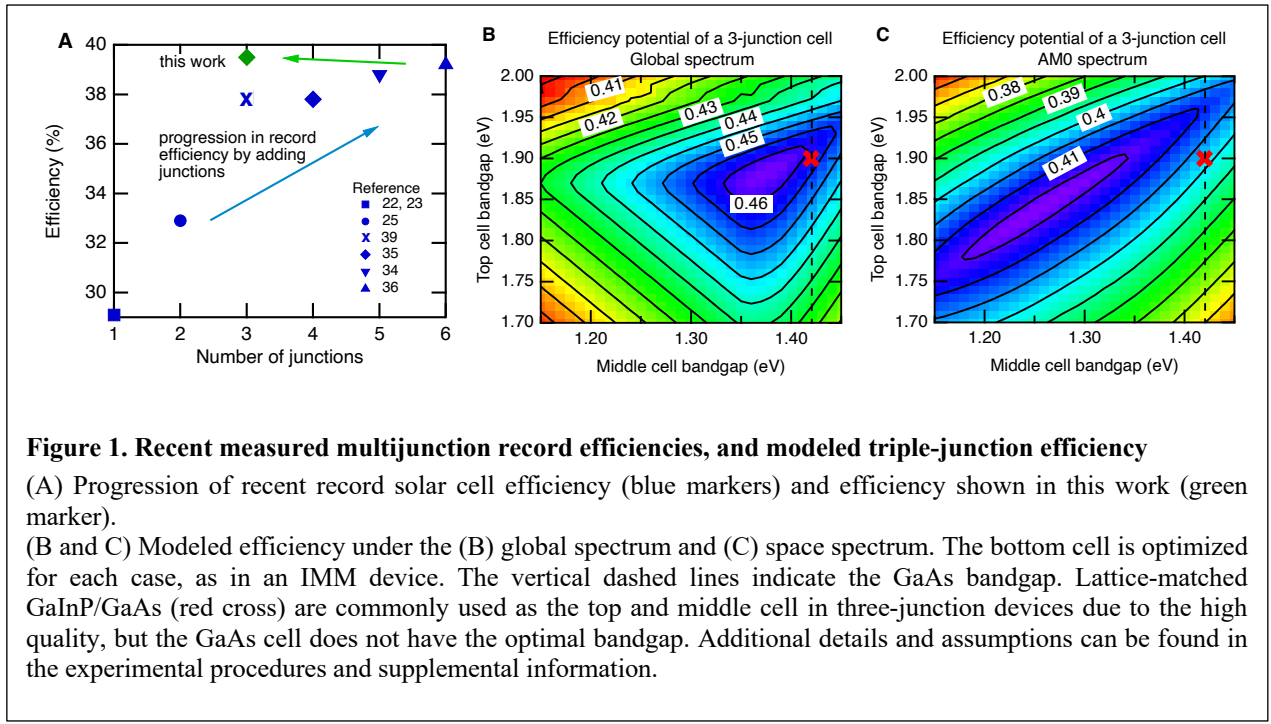

**Figure 1. Recent measured multijunction record efficiencies, and modeled triple-junction efficiency**
(A) Progression of recent record solar cell efficiency (blue markers) and efficiency shown in this work (green marker).
(B and C) Modeled efficiency under the (B) global spectrum and (C) space spectrum. The bottom cell is optimized for each case, as in an IMM device. The vertical dashed lines indicate the GaAs bandgap. Lattice-matched GaInP/GaAs (red cross) are commonly used as the top and middle cell in three-junction devices due to the high quality, but the GaAs cell does not have the optimal bandgap. Additional details and assumptions can be found in the experimental procedures and supplemental information.

As an alternative to metamorphic growth, quantum wells (QWs) can also be used to engineer the bandgap of a material. QWs are planar nanostructures that sandwich a lower bandgap well layer between two higher bandgap confining barrier layers, with an absorption edge that depends on the low bandgap of the well layer plus the effects of strain and quantum confinement, which act to raise the effective bandgap.[43,44] Strained layers with thickness below the Matthews-Blakeslee critical thickness maintain coherency with the growth substrate,[45] constraining individual QW thickness to only ~10 nm, and so many QWs are needed in order to achieve significant absorption of the incident light. Compressively-strained GaInAs QWs can be strain-balanced using tensile-strained GaAsP barriers,[46] and 50-100 strain-balanced QWs have previously been demonstrated.[47] QWs thus allow small modifications to the absorption edge without the formation of dislocations, and so can have higher radiative efficiency than metamorphic material.[42] The small reduction in bandgap is appropriate for the middle cell bandgap of a three-junction device, enabling a potential efficiency improvement over existing devices.

In this paper, we first develop an optically-thick GaInAs/GaAsP strain-balanced quantum well solar cell by using thin GaAsP barriers, and investigate tradeoffs between the number of quantum wells and device performance. GaAsP barriers with increased tensile strain require less thickness to strain-balance the GaInAs wells, thus increasing the fraction of the intrinsic region that is made of well material and increasing absorption for a given intrinsic region thickness. Here, we show devices with thin-barrier QWs that exhibit excellent absorption and voltage, a feat not demonstrated in over 30 years of research on quantum well devices.

We then incorporate high performance quantum wells into a triple-junction inverted metamorphic solar cell with an optimal bandgap combination under the global spectrum and a near-optimal bandgap combination under the AM0 space spectrum. The triple-junction consists of a GaInP top cell, a GaInAs/GaAsP strain-balanced QW middle cell, and a lattice-mismatched GaInAs bottom cell with low threading dislocation density, each of which has been highly optimized over decades of research. The solar cell achieves 39.5% efficiency under the global spectrum, which exceeds that of the previous record six-junction cell, and 34.2% efficiency under the space spectrum.

## RESULTS AND DISCUSSION

### Optically-thick quantum well solar cells using thin barriers

Despite having beneficial properties, quantum well solar cells have not enabled efficiency increases until recently due to practical limitations.[25] Quantum wells are typically placed in the intrinsic (undoped) region of the device so that the carrier transport occurs by drift in the electric field.[48] Since the width of the intrinsic region is determined by the actual carrier concentration, the number of quantum wells is limited by the minimum achievable doping level, which in turn limits absorption. In addition, devices with thick intrinsic regions tend to have increased non-radiative recombination and so lower radiative efficiency. Most importantly, achieving low non-radiative recombination in a quantum well solar cell requires excellent epitaxial material, including abrupt interfaces, fine control over strain-balancing, and control over elastic deformation via thickness modulation.[49,50] These practical challenges have limited the performance achieved in most previous work.[44,46,51]

The total absorption in the GaInAs/GaAsP quantum well (QW) solar cell depends on the thickness and number of quantum wells. Recently, the authors developed strain-balanced GaInAs/GaAsP quantum well solar cells with excellent radiative efficiency[25] but with optically thin wells, relying on an external, reflective back contact to provide a second pass of light through the QWs and even so only achieving 60-70% external quantum efficiency in the QW wavelength range. In Ref. [25], 8.5 nm $Ga_{.89}In_{.11}As$ quantum wells were strain-balanced with 17 nm GaAs$_{.9}$P$_{.1}$ barriers, resulting in an intrinsic region that was comprised of 33% GaInAs and 67% GaAsP, and QWs with an absorption edge of 1.35 eV.

Strain-balancing GaInAs with highly strained and thinner GaAsP leads to higher fractional GaInAs content, and would increase the total absorption for a given fixed depletion width. Here, we design single-junction GaInAs/GaAsP strain-balanced quantum well solar cells using thinner 5 nm GaAs$_{.68}$P$_{.32}$ barriers with 8.5 nm Ga$_{.89}$In$_{.11}$As wells to increase absorption. The thicknesses and compositions satisfy the strain-balancing condition derived in Ref. [52] and are shown in Fig. S4. We verified that the net average stress in the QWs was nearly zero by measuring the wafer curvature with an optical probe during growth. Growth conditions and sequences were chosen to result in abrupt interfaces and limited lateral thickness variation. Additional information on MQW design and development is described in the supplemental experimental procedures and Figures S2-S4.

Figure 2 shows the final triple-junction structure where these MQWs are intended. The single junction test structure is identical to the middle cell of the three-junction device, with contact layers outside of the GaInP barrier layers. An inverted rear-heterojunction structure was used, incorporating a 1-μm, silicon-doped n-type GaAs emitter and a 0.3-μm, zinc-doped p-type GaInP base. Details of the MQW structure are shown on the right of Fig. 2. The QWs were undoped and situated between the n-GaAs and p-GaInP, with 50 nm undoped GaAs on either side to minimize the impact of potential diffusion. The structure was grown in an inverted configuration due to its intended use in an inverted metamorphic multijunction, and was processed with a reflective back contact behind the quantum well solar cells that increases absorption, though there will not be such a reflector in the final triple-junction. The number of quantum wells was varied systematically from 100 to 300, which increases the cumulative GaInAs well-layer thickness from 850 nm to 2.5 μm, respectively. Further description is included in the experimental procedures.

Figure 3A compares the EQE of the quantum well devices to that of a baseline rear-heterojunction GaAs solar cell. The EQE of the quantum well cells at 800 nm is equivalent to that of the GaAs cell, confirming excellent carrier collection in all devices. The effective bandgap shift between the GaAs cell (880 nm) and the QW cells (925 nm) is clearly observed. The solar cell with 100 QWs has a higher EQE than previous work due to the increased thickness of GaInAs in addition to the extra absorption provided by the reflector.[25,47] The reflector also turns the cell into a Fabry-Perot cavity, where forward-travelling and backward-travelling waves interfere coherently and lead to the small oscillations that can be observed in the EQE. As the total thickness of GaInAs is increased, the EQE in the long wavelength range increases and the oscillations decrease as the absorption approaches unity. The device with 200 wells is optically thick due to the presence of the reflector, while the device with 300 wells has over 2.5 μm of GaInAs and so would be optically thick even without the reflector. Interestingly, there is a small bandgap shift as the number of wells increases, implying a minor material change that could be related to the gradual onset of lateral thickness or composition modulation.[49,50,53] Compared to the GaAs baseline EQE, the cells with 300 QWs gain 2.6 mA/cm$^2$ current under the global spectrum, and 3.5 mA/cm$^2$ under the AM0 space spectrum.

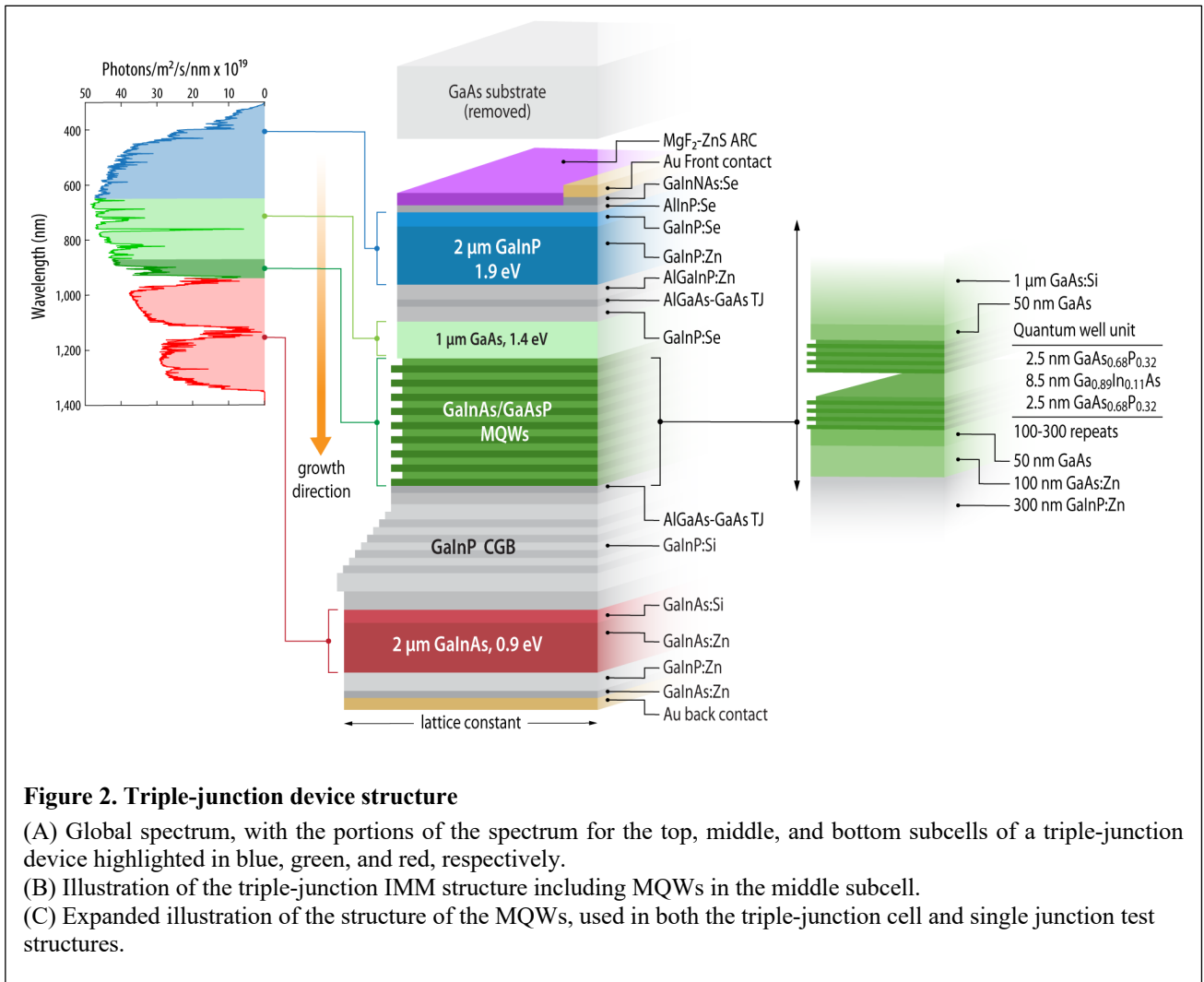

**Figure 2. Triple-junction device structure**
(A) Global spectrum, with the portions of the spectrum for the top, middle, and bottom subcells of a triple-junction device highlighted in blue, green, and red, respectively.
(B) Illustration of the triple-junction IMM structure including MQWs in the middle subcell.
(C) Expanded illustration of the structure of the MQWs, used in both the triple-junction cell and single junction test structures.

Figure 3B shows the dark J-V of the quantum well cells, compared to GaAs reference devices. One GaAs device (black line) was doped as a standard rear-heterojunction device, while another GaAs device was a p-i-n device (gray line) with a large intrinsic region, like the QW devices. The p-i-n GaAs cell is dominated by space-charge recombination at all measured voltages (parameterized by a J-V dependence of $J \sim J_{02} e^{qV/2kT}$ where $J_{02}$ is the saturation current density for SRH recombination), due to the thick depletion region, while the p-n device has less $J_{02}$ recombination and transitions to quasi-neutral recombination at around 1 V (parameterized by a J-V dependence of $J \sim J_{01} e^{qV/kT}$ where $J_{01}$ is the saturation current density in the drift-diffusion model). The QW devices show similar behavior to the p-i-n baseline cells.

One-sun illuminated J-V curves are shown in Figure 3C along with performance metrics. Devices without anti-reflection coating (ARC) are shown and analyzed, to avoid the influence of the ARC on the shunt resistance. The open-circuit voltage (Voc) of the 100 QW device is 1.03 V. The difference in Voc between the p-n GaAs cell and the QW solar cells is only 60 mV, roughly equivalent to the bandgap change. However, the fill factor (FF) of the p-n GaAs device is 86% while the p-i-n GaAs and 100 QW devices (which have identical intrinsic region thickness) have a FF of 82%, the consequence of increased $J_{02}$ recombination in p-i-n devices. As the intrinsic region thickness increases with additional QWs, the FF drops to 80% in the 300 QW device in part due to increased depletion region recombination.

These QW devices depend on carrier collection via drift in the electric field of the depletion region, and so may have a voltage-dependent carrier collection. The influence of the depletion width on carrier collection was determined by comparing the FF of the illuminated J-V with the FF calculated from the superposition of the photocurrent and the dark J-V. All samples had an insignificant <0.2% difference between those fill factor calculations, and no trends in FF were observed with the number of wells, indicating that carrier collection is nearly complete even at the maximum power point. In addition, the carrier concentration was measured using a standard capacitance-voltage (C-V) technique,[54] shown in Fig. 3D. The 300 QW sample has a background p-type doping as low as 2 x $10^{13}$ cm$^{-3}$. The apparent doping increases near the rear of the depletion region can be attributed to depth resolution limitations of the C-V measurement from the Debye length of low-doped samples.[55] The low doping results in a fully depleted QW region, and so agrees with the EQE and FF calculations.

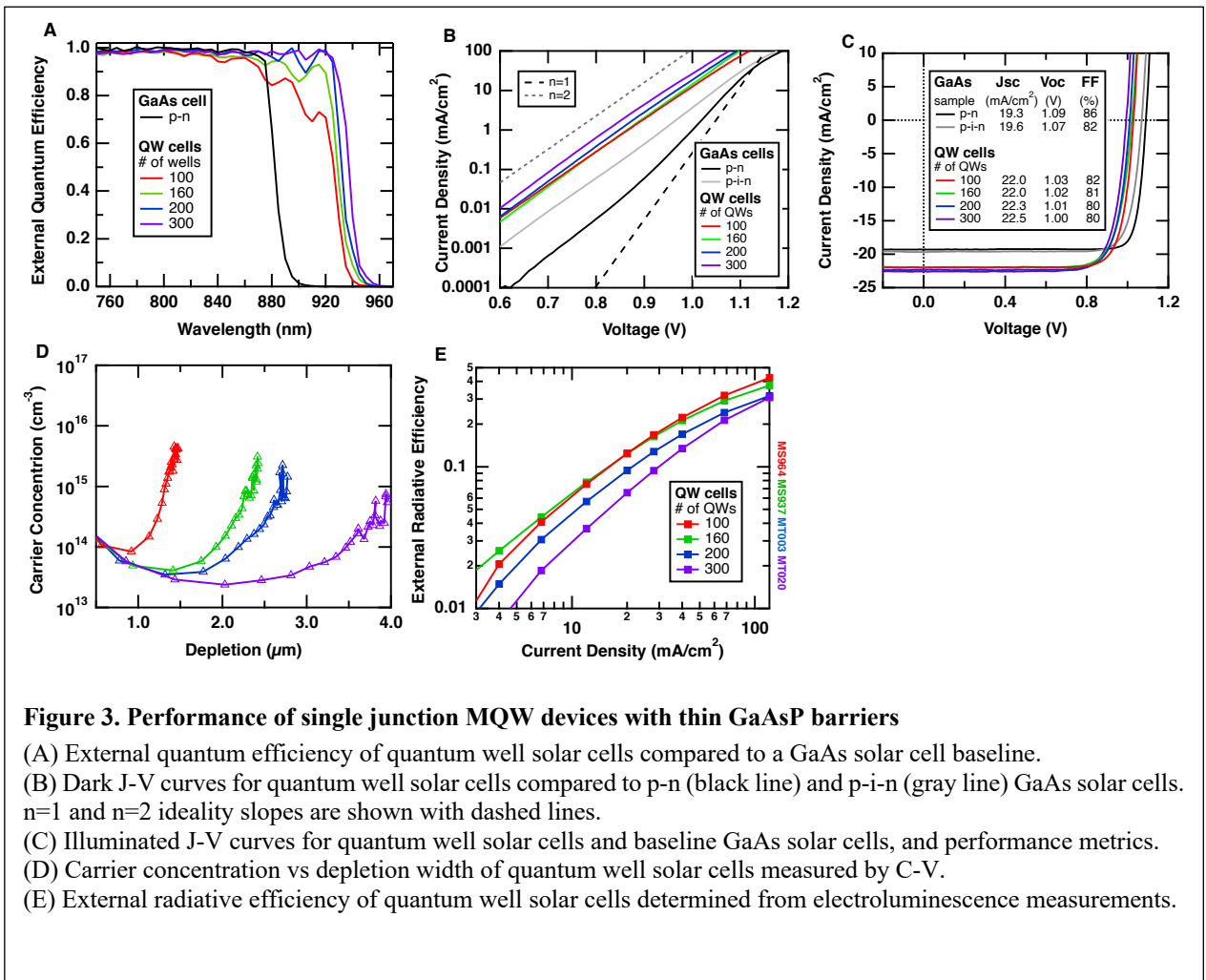

**Figure 3. Performance of single junction MQW devices with thin GaAsP barriers**
(A) External quantum efficiency of quantum well solar cells compared to a GaAs solar cell baseline.
(B) Dark J-V curves for quantum well solar cells compared to p-n (black line) and p-i-n (gray line) GaAs solar cells. n=1 and n=2 ideality slopes are shown with dashed lines.
(C) Illuminated J-V curves for quantum well solar cells and baseline GaAs solar cells, and performance metrics.
(D) Carrier concentration vs depletion width of quantum well solar cells measured by C-V.
(E) External radiative efficiency of quantum well solar cells determined from electroluminescence measurements.

There is a 30 mV loss in voltage as the number of quantum wells increases from 100 to 300. However, there is also a small bandgap change that can explain part of the voltage loss. To separate the impact of bandgap from other losses, electroluminescence was used to calculate the external radiative efficiency, plotted in Fig. 3E. All devices have excellent radiative efficiency, with the 100 QW device having over 40% radiative efficiency at 100 mA/cm$^2$. While the 100 and 160 QW samples have nearly identical performance, there is a small loss in ERE in the 200 and 300 QW samples, indicating that the loss in voltage in these samples is greater than the change in bandgap. The loss could be related to additional depletion region recombination or a material degradation such as lateral thickness or composition modulation, but the loss is minor since even the 300 QW sample has an ERE of 10 % at 30 mA/cm$^2$.

**Triple-junction inverted metamorphic devices with quantum wells**

Quantum wells were incorporated into the middle cell of a triple-junction inverted metamorphic solar cell. The top cell is a front homojunction GaInP,[56] and the bottom cell is a high performance metamorphic GaInAs cell accessed with a transparent GaInP compositionally graded buffer,[35,38] both with excellent material quality. The cells were interconnected using tunnel junctions described previously,[57,58] and Fig. 2 shows the full structure. The middle cell has 184 GaInAs wells to enable the appropriate photocurrent and is otherwise identical to the devices in the above study. Devices were tuned for the global and AM0 spectra by modifying the thickness of the top cell and changing the bandgap of the bottom cell by adjusting the graded buffer layer and the composition of the GaInAs.[35] Strain-balancing, relaxation monitoring, and material analysis was performed by in-situ wafer curvature and ex-situ X-ray diffraction (XRD), shown in Figure S5 of the supplemental information.

Figure 4A shows the EQE of the triple-junction global device, overlaid on the AM1.5g spectrum. The EQE in the QW region is over 80%, and the bandgap extends to the water absorption gap in the solar spectrum. The absorption in the cell is high enough to provide an appropriate photocurrent to the multijunction without the use of an internal reflector.[59] The summed internal quantum efficiency (IQE) is excellent in the range of the QWs, indicating a low loss of carriers associated with the QWs. A slight loss at shorter wavelengths (650 – 800 nm) highlights some absorption in the top tunnel junction, equivalent to a loss of <0.3 mA/cm$^2$. The integrated cell photocurrents are shown in Fig. 4A, along with the bandgaps. The top cell has the lowest photocurrent, and the middle and bottom cells have excess current that increases the fill factor of the multijunction device. Though it is often thought that an ideal multijunction is exactly current-matched, in actuality some excess current in lower cells is desirable in order to accommodate the lower fill factors and to maximize the multijunction current at the maximum power point.[60]

The EQE of the triple-junction space device is shown in Fig. 4B, overlaid on the AM0 spectrum. Comparing Fig. 4A and 4B highlights important differences between global and AM0 designs. While the QWs significantly increase the photocurrent in the middle cell compared to a GaAs baseline, the middle cell bandgap is still higher than the optimum shown in Fig. 1 for the AM0 spectrum. For a middle cell bandgap of 1.35 eV, the top cell bandgap is lower than optimal due to the use of GaInP rather than AlGaInP, which can have additional nonradiative recombination.[61] These two factors guide the AM0 design of this cell. With respect to the global design, the top cell was thinned from 2 µm to 1 µm in the space design to accommodate the slightly non-optimal top cell bandgap and boost current to the middle cell. The bottom cell bandgap is tunable by changing the graded buffer, and was raised in order to boost the voltage. Changes in the top and middle cell bandgaps were due to unintentional reactor drift. The Herpin MgF$_2$/ZnS ARC was designed to maximize current to the top two cells.[62] An even lower bandgap middle cell would enable a more optimal bandgap combination under the space spectrum, as would an optically thick GaInP top cell.

Electroluminescence was used to determine the dark J-V curves of each cell, shown in Fig. 4C. The electroluminescence spectra and external radiative efficiency are shown in Fig. S7. The voltages from the global 3J at 15 mA/cm$^2$ current were determined for the top, middle, and bottom cells to be 1.47, 0.98, and 0.55 V, respectively, resulting in bandgap-voltage offsets $W_{oc}$ (=$E_g/q - V_{oc}$) of 0.41, 0.35, and 0.35 V, respectively. Note that the subcell voltages sum to 20 mV higher than the measured device voltage (Fig. 4B) due to rounding and measurement error. The material quality in all cells has been highly optimized resulting in excellent cell voltages. The top cell is a high performance front-junction

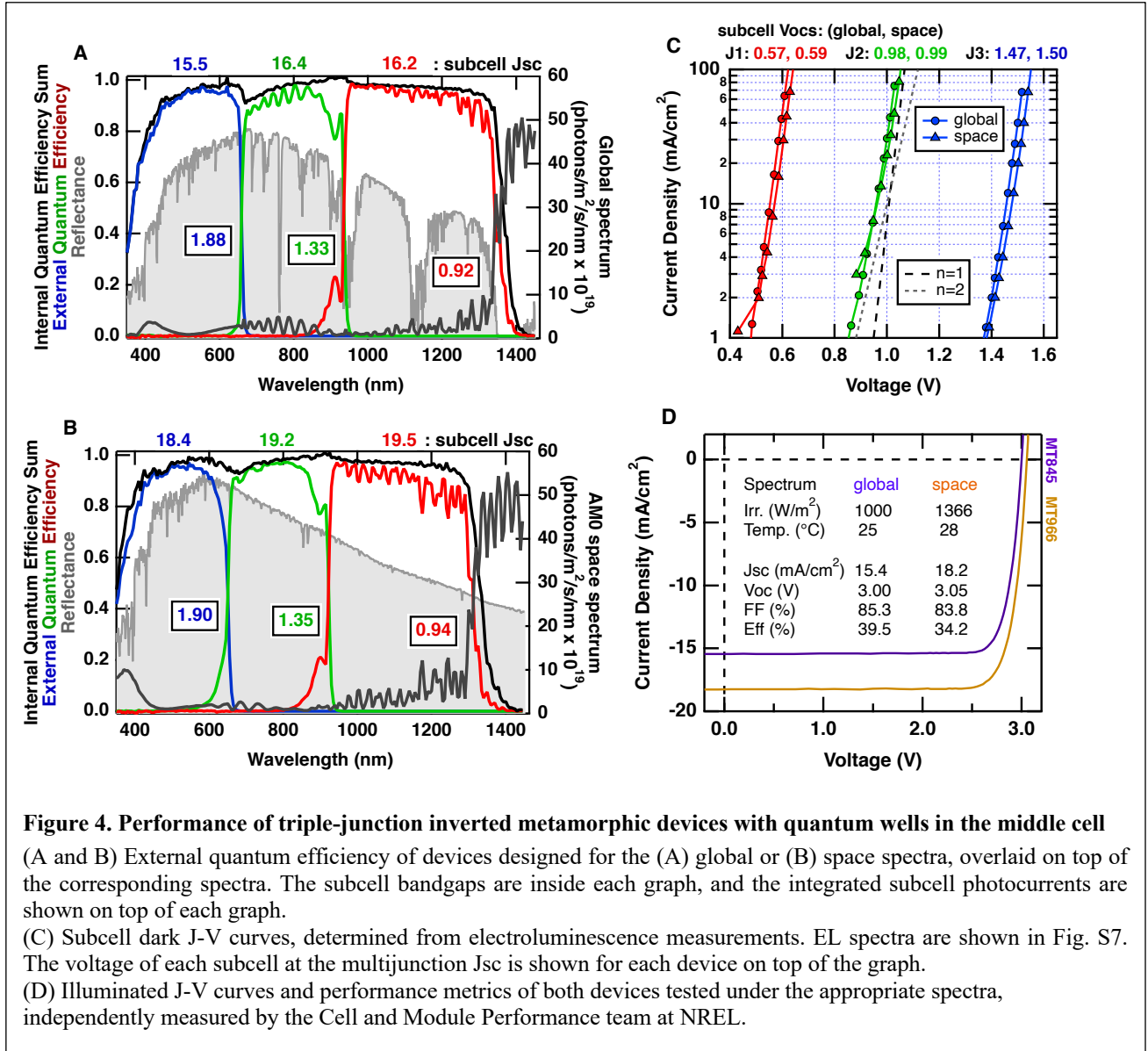

**Figure 4. Performance of triple-junction inverted metamorphic devices with quantum wells in the middle cell**
(A and B) External quantum efficiency of devices designed for the (A) global or (B) space spectra, overlaid on top of the corresponding spectra. The subcell bandgaps are inside each graph, and the integrated subcell photocurrents are shown on top of each graph.
(C) Subcell dark J-V curves, determined from electroluminescence measurements. EL spectra are shown in Fig. S7. The voltage of each subcell at the multijunction Jsc is shown for each device on top of the graph.
(D) Illuminated J-V curves and performance metrics of both devices tested under the appropriate spectra, independently measured by the Cell and Module Performance team at NREL.

GaInP cell described in Ref. [56], and the bottom cell has one of the lowest reported $W_{oc}$s for a metamorphic cell.[63] The QW cell maintains a low $W_{oc}$ within the multijunction despite the lack of a back reflector. The QW cell has an n=2 diode characteristic at low currents, and begins transitioning to

n=1 at 10 mA/cm$^2$. The voltages for the space device are also shown in Fig. 4D, and are similar with respect to the subcell bandgap except for a slightly lower FF and voltage for the QW cell.

The JV curves of the devices are shown in Fig. 4D, measured under both the global spectrum (at 25 °C and 1000 W/m$^2$ irradiance) and AM0 space spectrum (at 28 °C and 1366 W/m$^2$ irradiance). The supplemental experimental procedures contain measurement details and the measurement spectra is shown in Figure S1. Under the global spectrum, the device measures (39.5 ± 0.5)% efficiency, the highest one-sun efficiency solar cell of any type as of this writing. Under the AM0 space spectrum, the device measures (34.2 ± 0.6)%, the highest beginning-of-life triple-junction device yet reported under the AM0 spectrum, and close to that of a five-junction wafer bonded device.[34]

**Conclusions**

We have developed optically thick GaInAs/GaAsP strain-balanced quantum well superlattices with excellent absorption and voltage. The high absorption is a result of the very high fraction of quantum well material within the depleted region of the cell, and the high voltage is due to excellent strain-balancing, interface, and material control. Optically thick quantum well superlattices allow bandgap modification with minimal loss, and so enable multijunction solar cells with optimal bandgap combination. We implemented these quantum well solar cells into triple-junction inverted metamorphic multijunction solar cells, where they produce sufficient photocurrent for the multijunction device without the use of an internal reflector. The QW middle cell was combined with a GaInP top cell and a high-performance lattice-mismatched GaInAs bottom cell, where each cell material was highly optimized through sustained research into GaInP, GaInAs quantum wells, and lattice-mismatched GaInAs with low defect density. The optimal bandgap combination of the device, combined with the high voltage and high absorption in the QW cell, leads to a record (39.5 ± 0.5)% efficiency under the global spectrum and (34.2 ± 0.6)% efficiency under the AM0 space spectrum.

These triple-junction devices have higher efficiency than previous 1-sun global record devices that used 5 or 6 junctions even though they collect less of the incident solar spectrum and have fewer junctions. These cells are most useful for highly area-constrained terrestrial applications and for low radiation space missions where efficiency is critical, and indeed are closely related to commercial products such as the triple-junction IMM cell that has been integrated into high-altitude aerial vehicles.[64] If designed for the concentrated direct spectrum, the cells would likely outperform standard triple-junction IMM cells, making them good candidates for concentrator photovoltaics (CPV) applications. Additional research towards reducing the cost of III-V solar cells through substrate recovery and streamlined processing would increase the terrestrial market, and investigation of radiation hardness for improved end-of-life efficiency would increase the potential space applications[65-67]. As these are the highest efficiency 1-sun solar cells as of this writing, these cells also set a new standard for achievable efficiency across all PV technologies.

**EXPERIMENTAL PROCEDURES**

**Epitaxial growth**

All samples were grown by metalorganic vapor phase epitaxy (MOVPE) in a custom-built atmospheric pressure reactor on (001) GaAs substrates, miscut 2° towards (111)B. Standard III-V MOVPE precursors were used, including trimethylindium, triethylgallium, trimethylgallium, and trimethylaluminum as the group III sources, and arsine, phosphine, and dimethylhydrazine as the group V sources. Disilane, diethylzinc, carbon tetrachloride, and hydrogen selenide were the dopant precursors. All sources were mixed with a purified $H_2$ carrier gas flowing at 6 lpm. A detailed structure of the 3-junction device is illustrated in Figure 2 of the main text. The substrate was first etched for one minute in $NH_4OH : H_2O_2 : H_2O$ (2:1:10 by volume), rinsed in DI water, and loaded into the reactor under a $N_2$ overpressure. The growth began with a deoxidation at 700°C under an arsine overpressure, followed by a GaAs nucleation layer, then followed by GaInP stop etch for post-growth substrate removal. Then, the GaInNAs front contact layer was grown followed by the remaining layers in Fig. 2. The growth direction was inverted, meaning that the high bandgap subcells were grown prior to low bandgap subcells. Single-junction quantum well devices had an identical structure to the middle cell of the multijunction device. GaAs comparison devices were grown with identical thickness to the quantum well devices, except using doped or undoped GaAs in the MQW region rather than QWs. Growth temperatures between 620-700 °C were used for all layers except the tunnel junction layers and contact layers, which were grown at lower temperatures, 570-620 °C, to encourage dopant incorporation. The superlattices were grown at 650 °C. All layers employed growth rates between 2-7 µm/hr, and the superlattice layers were grown between 2.5 – 3.1 µm/hr without stop-growths between layers.

**Device processing**

Both single-junction and triple-junction devices were processed using an inverted processing sequence described previously.[68] Briefly, the surface of the sample was first electroplated with planar Au, which acts as both a back contact and rear reflector. Then, the sample was inverted and bonded onto a Si handle using a low-viscosity epoxy, and the substrate was removed using $NH_4OH: H_2O_2$ (1:3 by volume) chemical etchants. The front contact layer was exposed and Ni/Au front contacts were electroplated using standard photolithography techniques,[69] and then the front contact layer was etched away using the metal contacts as a mask. Devices were then isolated using selective chemical etchants, and finally a $MgF_2$/ZnS anti-reflection coating (ARC) was deposited by thermal evaporation. The ARC deposition for both the global and space cells begins with a very thin, ~1 nm, layer of $MgF_2$ to promote adhesion of the dielectrics to the semiconductor. Then, 50 nm ZnS, 15 nm $MgF_2$, 15 nm ZnS, and 100 nm $MgF_2$ are deposited for the global cell, a four-layer coating that is a Herpin-equivalent of a three-layer structure.[70] The ARC for the AM0 cell is similar but all layers are slightly thinner to promote current in the top two cells: 45 nm ZnS, 12 nm $MgF_2$, 12 nm ZnS, 90 nm $MgF_2$. The reflection of these ARCs is shown in Fig. 4.

**Efficiency Modeling**

Multijunction cell efficiency is modeled using a 1-D model using realistic Wocs (Eg/q – Voc) at 16 mA/cm$^2$ of 0.38 V, 0.36 V, and 0.34 V for the top, middle, and bottom cell, respectively, to calculate the dark current for each junction through the standard cell equation Voc ≈ ($nkT/q$) ln(Jsc /Jo). Jo is modified for temperature using Jo (T) = const $*$ T$^3$ exp(−Eg (T)/kT). The quantum efficiency (QE) of each junction is modeled using absorption coefficients of GaInP for the top junction and GaAs for the middle and bottom junctions, with the curve shifted appropriately for the bottom cell. Absorption loss in the AlInP window was included using absorption coefficients and nominal thickness. The absorption coefficients are shifted due to temperature by using Varshni parameters from the literature. All bandgaps in Figure 2 are at 300 K. The efficiency was determined as a function of the top and middle subcell bandgaps, allowing for subcell thinning in each case. The bottom cell bandgap was allowed to vary and was optimized for each combination of the top and middle subcell bandgaps. These constraints were chosen because of the intent of a three-junction inverted metamorphic device, where the top and middle subcell bandgaps are constrained by the substrate lattice constant, but the bottom cell is metamorphic and thus the bandgap is less constrained.

**External quantum efficiency**

External quantum efficiency (EQE) and reflection were measured on a custom-built tool designed for multijunction analysis. Devices were illuminated with 315 Hz chopped, monochromated light from 350 nm to 1450 nm while being continuously illuminated with light emitting diodes (LEDs). 470 nm and 740 nm LEDs were used when measuring the bottom cell, 470 nm and 1200 nm LEDs were used when measuring the middle cell, and 740 nm and 1200 nm LEDs were used when measuring the top cell. No voltage biasing was needed for any subcells except the bottom junction of the global device, where 2.4 V bias was used. A low-noise current-voltage preamplifier and an SR830 lock-in amplifier were used to extract the device current, which was compared to a calibrated reference cell. EQE was corrected for luminescent coupling using the procedure described in Ref. [71]. Near-normal specular reflection was measured on the same tool using a stacked Si/Ge photodiode.

**I-V measurement**

I-V curves were acquired for single junction devices and triple junction devices using different tools and techniques. The three-junction devices are record efficiency devices, and their I-V data was independently measured and certified by the Cell and Module Performance (CMP) team at NREL. The CMP team is one of the few PV testing labs in the world that is capable of calibrating primary reference cells, secondary reference cells/modules, single-junction and multijunction cells and modules. Both simulators are spectrally adjustable, and both utilize the EQE along with spectral mismatch factors to determine the irradiance. The triple-junction devices were measured with a photocurrent ratio of 1.0 between each junction pair, with less than 1% error in the simulator spectrum for each junction. Additional details are included in the supplementary experimental procedures.

**Capacitance-voltage**

Capacitance-voltage (C-V) was measured using an LCR meter using 3 kHz AC frequency and up to 3 V of DC reverse bias voltage. Carrier concentration and depletion width were determined using standard analysis techniques.[54] The low doping in the sample results in a large Debye length, which decreases depth resolution from the C-V measurement, but no modifications to the data were performed.[55,72]

**Electroluminescence**

Electroluminescence (EL) was measured at multiple current injections while collecting and analyzing light via a fiber optic and Spectral Evolution spectroradiometer. The tool was calibrated and light output was quantified using a technique described previously,[24,73] enabling determination of the external radiative efficiency and individual cell JV characteristics through the reciprocity theorem.[74]

## ACKNOWLEDGMENTS

The authors kindly thank Chuck Mack and Rafell Williams for measurement support, Alfred Hicks for illustration, Daniel Friedman, Kunal Mukherjee, Jeronimo Buencuerpo, and Meadow Bradsby for valuable conversations, and Ned Ekins-Daukes and colleagues for valuable conversations and foundational work. This work was authored by the National Renewable Energy Laboratory, operated by alliance for Sustainable Energy, LLC, for the U.S. Department of Energy (DOE) under Contract No. DE-AC36-08GO28308. Funding was provided by the U.S. Department of Energy Office of Energy Efficiency and Renewable Energy Solar Energy Technologies Office, under Award No. 34358. The views expressed in the article do not necessarily represent the views of the DOE or the U.S. Government. The U.S. Government retains and the publisher, by accepting the article for publication, acknowledges that the U.S. Government retains a nonexclusive, paid-up, irrevocable, worldwide license to publish or reproduce the published form of this work, or allow others to do so, for U.S. Government purposes.

## SUPPLEMENTAL EXPERIMENTAL PROCEDURES

### Single junction I-V measurement

Illuminated and dark current-voltage curves from single junction devices were measured using a custom-built simulator using a XT10 xenon bulb as a light source. A calibrated reference cell, calibrated in 2020 by the CMP team at NREL, was used with the device quantum efficiencies to determine spectral mismatch factors for each cell and set the light intensity.[1] The simulator spectrum for single junction cells is shown in Fig. S1. The single junction devices were measured in air in standard laboratory conditions, at a stage temperature of 25 °C. No light soaking or cell preconditioning was used. The device area, 0.25 cm$^2$, was defined as the area of the photolithography mask for mesa isolation, and the front contact busbar is included in the device area. The I-V scans were acquired from forward bias to reverse bias, using at least 4 ms dwell time at each voltage. Representative cell data are shown in Figure 3.

### Triple junction I-V measurement

Illuminated current-voltage for the triple-junction devices was independently measured and certified by the Cell and Module Performance (CMP) Team at NREL using a one-sun multi-source simulator. The simulator consists of 9 adjustable spectral channels split from two 1500-W Xe (Cermax PE1500D-13FM Xenon Arc lamp) and two 750-W tungsten (OSRAM GLD 750W/115V) lamps. The light from each channel was coupled to a light integrator box (homogenizer) via large optical fiber bundles, and the light intensity was adjustable through variable apertures at the entrance of each fiber bundle.

The photocurrent ratio for triple-junction devices was determined by using the measured spectrum and refence spectrum using Ref. [2]. The spectrum was measured using an annually calibrated ASD FieldSpec 3 spectroradiometer, calibrated on April 23$^{rd}$, 2021, over the range of 300 – 2400 nm, and modified so that the effective junction current ratio, $R_{ij}$, is 1.00 with less than 1% error in each junction. The junction current ratios are defined based on the measured spectra:

$$R_{ij} = \frac{\int \Phi_{Ref}(\lambda)Q_i(\lambda)d\lambda}{\int \Phi_{Ref}(\lambda)Q_j(\lambda)d\lambda} \times \frac{\int \Phi_{Sim}(\lambda)Q_j(\lambda)d\lambda}{\int \Phi_{Sim}(\lambda)Q_i(\lambda)d\lambda} \qquad \text{(Equation S1)}$$

where $\Phi_{Ref}$ is the reference spectrum, $\Phi_{Sim}$ is the measured simulator spectrum, and $Q_i$ is the EQE of the ith junction. Accordingly, a spectrum-adjustment algorithm that relies solely on the QEs of the subcells was developed to adjust the simulator spectrum until it meets the desired reference spectrum condition.[2] The absolute irradiance was then adjusted to 1000 W/m$^2$ for AM1.5G spectrum standard, or 1366.1 W/m$^2$ for the AM0 spectrum standard, by changing the distance to an annually calibrated silicon reference cell, calibrated on May 24$^{th}$, 2021. The spectrum for each triple-junction cell is plotted in Fig. S1, where the relative intensity was collected by the spectroradiometer and the absolute intensity was adjusted using a calibrated cell.

The stage temperature was 28 °C and 25 °C for the AM0 and AM1.5 cells, respectively, and the devices were measured in air under standard laboratory conditions. No light soaking or cell preconditioning was used. The I-V scan consisted of 60 I-V pairs starting at 101% of Voc (forward bias) and sweeping to -

20% of Voc (reverse bias). Dwell time is approximately 100 ms at each bias voltage. Please note that the voltage increments are not uniform, with a denser voltage spacing in region of maximum power point. The illuminated area was defined by a double-layer metal aperture, and the aperture area, 0.242 cm$^2$, was measured under a calibrated Nikon microscope using the CMP team's standard protocol for area measurements. Record cell data are shown in Figure 4. The front contact busbar was included in the device area.

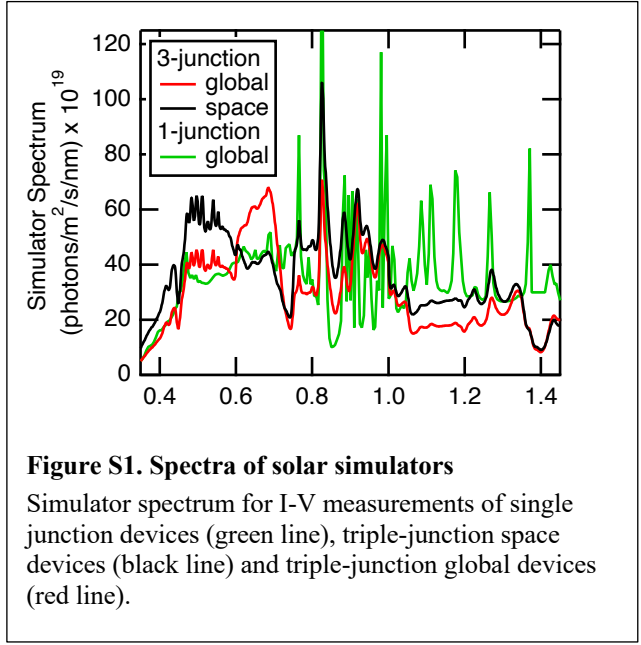

**Figure S1. Spectra of solar simulators**

Simulator spectrum for I-V measurements of single junction devices (green line), triple-junction space devices (black line) and triple-junction global devices (red line).

**X-ray diffraction**

X-ray diffraction (XRD) was measured on a Panalytical X'Pert Pro diffractometer using a 4-bounce (004) Ge crystal monochromator and a Pixel array detector. For superlattice modeling, single axis Omega-2Theta scans were acquired with a 55 μm electronic slit. Reciprocal space maps of (004) and (115)GE planes were collected using the array detector in 1D mode, and data was analyzed for tilt, strain, and composition.

**Wafer Curvature**

During growth, a 2-dimensional k-Space multi-beam optical sensor (MOS) monitored the wafer curvature using the deflection of a 3 x 3 parallel array of 662 nm laser beams. The optical path was near-normal, the laser array was aligned with the [110] and [-110] crystal axes, and no substrate rotation was used during growth. Average stress in the MQW layers and metamorphic bottom cell was determined using the Stoney equation[3,4] to ensure proper strain-balancing in the MQW layers and no plastic relaxation in either the MQW layers or metamorphic bottom cell. Importantly, MOS integrates the total stress of the MQW stack to determine overall strain-balancing in a complex MQW stack, including any unintentional interfacial layers, and so was the primary technique used for strain-balancing in the MQW layers and growth of strain-free metamorphic layers.

**MQW design**

The design of quantum well solar cells is complicated by both the strain and quantum confinement of the layers as well as material properties such as interface control, strain-balancing to prevent plastic relaxation, and impacts from elastic relaxation.

First, a bandgap target was determined by careful modeling of the optimal bandgap combination of a three-junction device under the global spectrum, combined with realistic considerations. As seen in Fig. 1 of the main text, the ideal bandgap for the middle cell under the global spectrum is about 1.36 eV. We anticipated that absorption in the QW region would not be complete, so initially targeted a slightly lower bandgap of 1.33 – 1.34 eV in order to maintain current matching, and this bandgap coincides with the absorption gap in the terrestrial solar spectrum at 925 nm.

GaInAs and GaAsP were chosen for the well and barrier material, respectively, due to their extensive history in QW systems, the appropriate bandgap and strain sense, as well as their ease of growth. Interface quality and susceptibility to elastic relaxation are both important practical considerations in material choice. Transitions between layers should not result in highly strained material or in low bandgap quaternaries. In the OMVPE reactor used in this study, abrupt interfaces are facilitated between GaInAs and GaAsP by establishing precursor flows prior to run/vent valve switching, and using a moderate temperature and high growth rate to limit group V intermixing.[5] Elastic relaxation through lateral thickness modulation or composition modulation has been observed previously in III-V QWs, and becomes increasingly problematic with increasing strain.[6-8] Here, we limit elastic relaxation by using high V/III ratio, moderate temperature, high growth rates, and low substrate miscut, described in the Epitaxial growth section.

The effective bandgap in quantum confined materials depends on the natural bandgap of the well material, the thickness and strain of the well, and, to lesser extent, the bandgap and thickness of the barrier.[9,10] The effects of strain are determined by the distortion of the unit cell and are calculated with knowledge of the elastic constants in the well and barrier materials. In general, strain separates the heavy hole and light hole bands in the valence band, and changes the position of the conduction band. For the QWs used here, the compressive strain in the well raises the energy of the heavy hole band and raises the energy of the conduction band, while the tensile strain in the barrier raises the energy of the light hole band and lowers the energy of the conduction band. The shifted energy levels due to quantum confinement are then found by solving the one-dimensional Schrödinger equation in a finite potential, for the strain-modified conduction and valence bands separately. At the voltages used in these cells, field-related distortions of the confining potential can be neglected. More details can be found in Ref. [5].

The thicknesses and compositions of the well and barrier materials are related by the need to strain-balance, limiting the degrees of freedom.[11] The relationship between these variables is given by

$$a_0 = \frac{A_w t_w a_w a_b^2 + A_b t_b a_b a_w^2}{A_w t_w a_b^2 + A_b t_b a_w^2} \quad \text{(Equation S2)}$$

where $a_0$, $a_w$ and $a_b$ are the relaxed lattice constants of the substrate, well and barrier layer materials, and $t_w$ and $t_b$ are the thicknesses of the well and barrier layers. $A_w$ and $A_b$ are biaxial elastic constants for the well and barrier materials and are in turn related to the stiffness coefficients $C_{11}$ and $C_{12}$ for each material through the relationship $A = C_{11} + C_{12} - 2C_{12}^2/C_{11}$. The elastic constants for the ternary alloys are calculated by interpolating from the endpoint binaries GaAs and GaP for the barrier, and GaAs and InAs for the well.

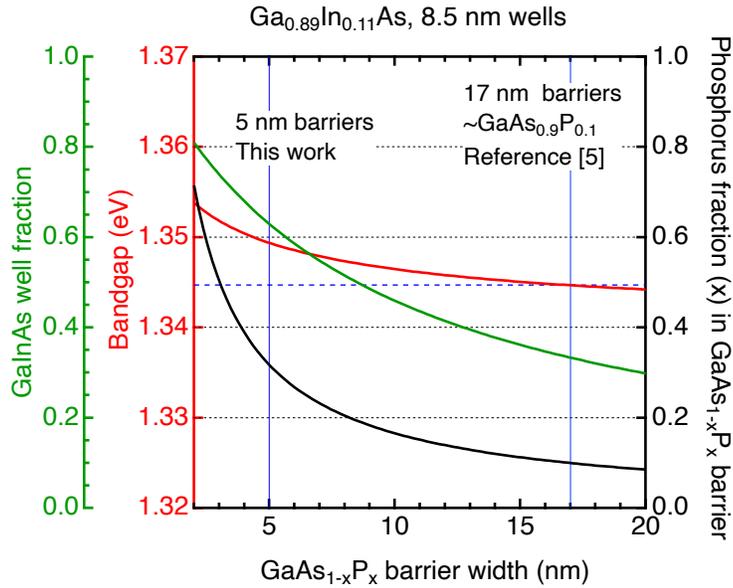

**Figure S2. Evolution of GaInAs quantum well bandgap with GaAsP barrier thickness**
The curves shown are the phosphorus composition of the barrier (black), the effective bandgap of the well (red), and the fraction of QW structure that is made of GaInAs (green). The well is $Ga_{0.89}In_{0.11}As$ and 85 Å for all points.

As the well thickness increases, the bandgap of quantum confined material decreases, which is beneficial for these devices. However, outside of dilute nitrides, III-V material with lower bandgap than GaAs is compressively strained on a GaAs substrate, and so the maximum thickness of the well is limited by the critical thickness of the material.[12] Previous work has shown the relationship between GaInAs composition and thickness and the effective bandgap.[9,13] While a range of thickness and composition combinations can target 1.33 eV, we use a combination that results in about half the critical thickness, $Ga_{0.89}In_{0.11}As$ and 85 Å, respectively.[14] Because the GaInAs thickness is well below the critical thickness, small changes to the GaInAs thickness and composition are possible without concern. When the 85 Å $Ga_{0.89}In_{0.11}As$ well is strain-balanced using GaAsP, the GaAsP barrier will be below its critical thickness for all possible strain-balanced thickness and compositions, even considering potential atomic ordering.[15]

In this study, the barrier thickness is thinned to enable a higher fraction of GaInAs within the MQWs. The impact of barrier thickness on the GaInAs QW bandgap is shown in Fig. S2. This model considers the strain and quantum confinement but assumes that the individual wells are still separate with localized wavefunctions, and therefore neglects the effects of any overlap of the wavefunctions due to tunneling. Since the well thickness and composition are assumed to be fixed, the model first applies the strain-balancing condition (EQ S2) at each barrier thickness to calculate the appropriate GaAsP composition, and then calculates the energy levels using the Schrödinger equation and elastic constants.

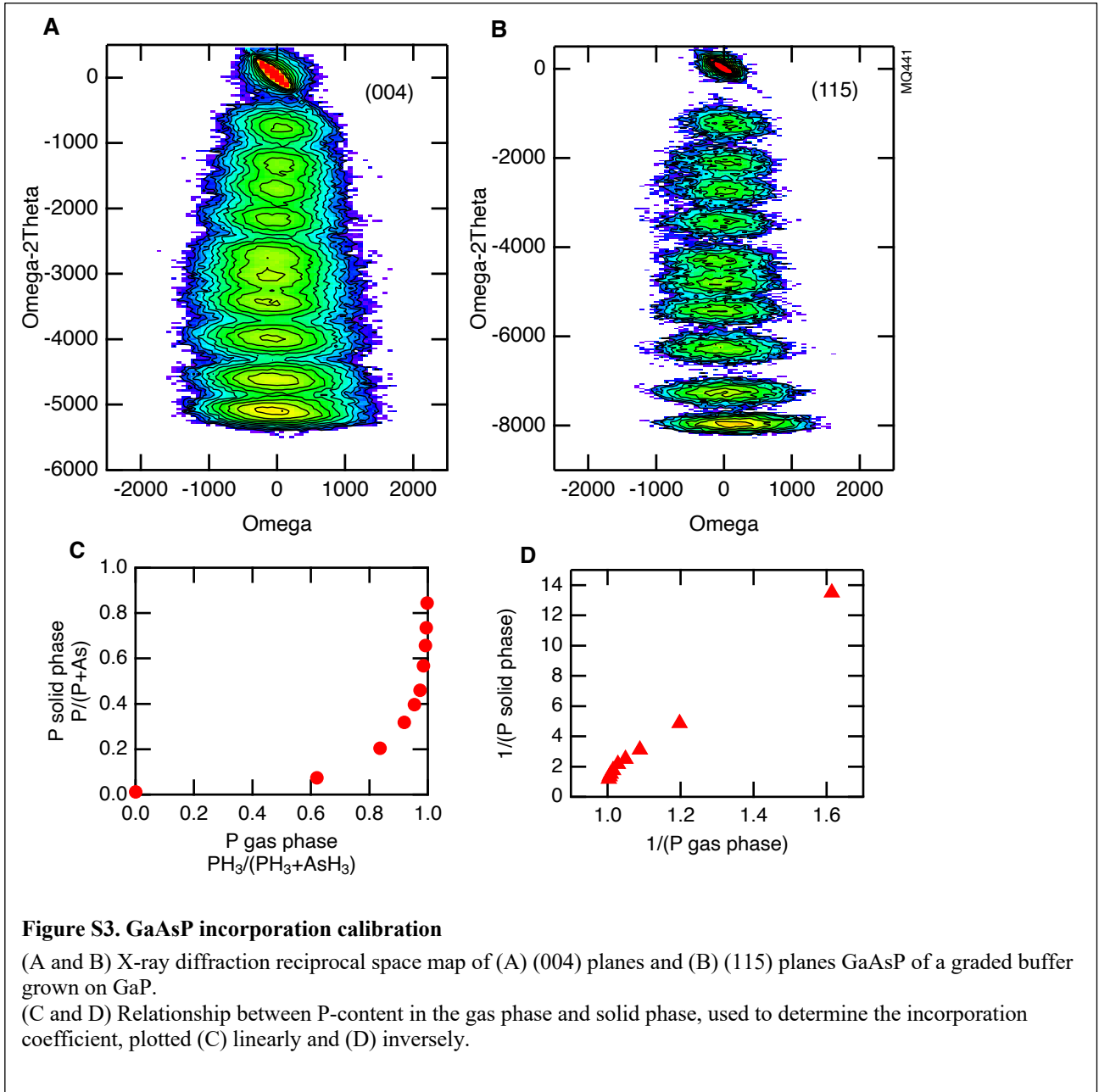

**Figure S3. GaAsP incorporation calibration**
(A and B) X-ray diffraction reciprocal space map of (A) (004) planes and (B) (115) planes GaAsP of a graded buffer grown on GaP.
(C and D) Relationship between P-content in the gas phase and solid phase, used to determine the incorporation coefficient, plotted (C) linearly and (D) inversely.

As the barrier is thinned, the phosphorus composition in the barrier increases (black curve) and along with it the barrier height. The higher confining potential leads to a slight increase in the QW bandgap (red curve), as determined by the lowest energy eigenstates in the conduction and valence bands. The green curve shows the increasing volume fraction of GaInAs in the MQW region as the barrier thins. 5 nm GaAsP thickness was chosen as a reasonably thin barrier without an exceptionally high strain in order to avoid elastic relaxation. The corresponding composition is $GaAs_{0.68}P_{0.32}$.

# MQW development

The $Ga_{0.89}In_{0.11}As$ composition was determined from XRD of a single-layer calibration run, and growth rate was determined from etch tests of GaInP and GaAs. The incorporation coefficient for GaAsP at the growth temperature (650 °C) was determined using a ten-layer GaAsP step-graded buffer on a (001) GaP substrate miscut 2° towards (111)B, where each step was 0.5 μm thick. X-ray diffraction of (004) and (115)GE planes to determine the composition, shown in Fig. S3. The incorporation of the mixed group V ternary is nonlinear and temperature dependent. Local incorporation coefficients around the desired GaAsP composition were utilized.

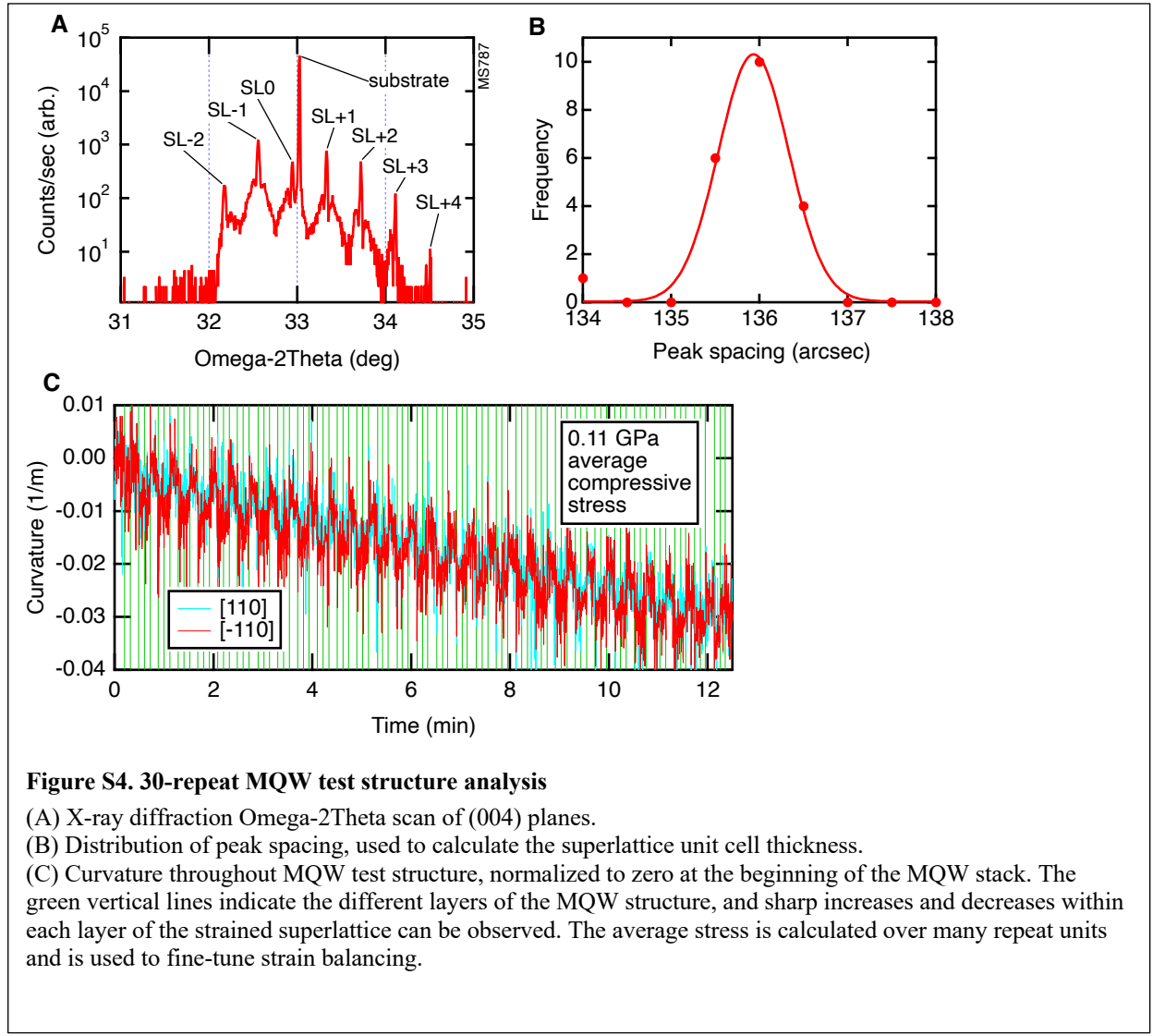

**Figure S4. 30-repeat MQW test structure analysis**
(A) X-ray diffraction Omega-2Theta scan of (004) planes.
(B) Distribution of peak spacing, used to calculate the superlattice unit cell thickness.
(C) Curvature throughout MQW test structure, normalized to zero at the beginning of the MQW stack. The green vertical lines indicate the different layers of the MQW structure, and sharp increases and decreases within each layer of the strained superlattice can be observed. The average stress is calculated over many repeat units and is used to fine-tune strain balancing.

Then, GaInAs and GaAsP layers, nominally strain-balanced, were implemented into a MQW test structure with 30 QW repeat units. XRD was used to confirm the superlattice thickness, shown in Fig. S4. Strong superlattice peaks are observed, with seven peaks that are distinct from the substrate. The superlattice unit cell thickness $\Lambda$ was estimated by applying the Bragg equation to the peak spacing:

$$\Lambda = \frac{(L_i - L_j)\lambda}{2(\sin \vartheta_i - \sin \vartheta_j)} \qquad \text{(Equation S3)}$$

where $L_i$ and $L_j$ denote satellite numbers (-2, -1, 0, …) for peaks at angles $\vartheta_i$ and $\vartheta_j$, and $\lambda=1.5046$ Å is the Cu-K$_{\alpha 1}$ x-ray wavelength. For the seven peaks, there are 21 combinations of spacing, and the distribution shows the peak to be $13.6 \pm 0.1$ nm, in close agreement with nominal values based off growth rate calculations. The superlattice peaks have some splitting which may be related to non-random interface roughness.[16,17]

Wafer curvature was used to measure strain-balancing. Wafer curvature can measure the average stress thickness of the entire stack and includes any impacts from unintentional interfacial layers, so that the structure is strain-balanced regardless of any deviations from nominal compositions. Fig. S4C shows the curvature of the MQW test structure with 30 repeat units, which showed an average compressive stress of 0.11 GPa. Feedback from wafer curvature was used to fine-tune the thickness and composition of the barrier to improve strain-balancing in the solar cell devices. Examination of interfaces and elastic relaxation was performed by TEM, discussed elsewhere.[5,18]

## SUPPLEMENTAL INFORMATION

### Triple-junction device strain

Strain relaxation should not occur in the triple-junction device except during the growth of the compositionally graded buffer, and so the alloy compositions of the many layers of this complex device need to be controlled. The thick layers of the device require extra care, since a small strain in a thick layer can easily exceed the critical thickness. Each subcell of the triple-junction device is >1 μm thick in order to absorb a significant portion of the incoming spectrum, making strain management in the GaInP top cell, the MQW region, and the GaInAs bottom cell critical. Wafer curvature, shown in Fig. S5, is used throughout the growth of the triple-junction device to ensure that each subcell is not overly strained, and that no strain relaxation occurs during the growth of the active subcells. Fig. S5 shows the curvature of the device designed for the space spectrum, and labels the location of the thick subcells. Although each subcell has some stress, observed by the non-zero slope of the curvature throughout the layer, there is no strain relaxation in these layers, which would be observed as a change in the slope of the curvature during these layers. Notably, the GaInAs cell has asymmetric curvature and thus stress between <110> directions, but the average curvature slope is low.

X-ray diffraction reciprocal space maps (RSMs) of the triple-junction device is also shown in Fig. S5. While wafer curvature gives a good indication of the strain throughout the device structure, XRD RSMs are also used to confirm that there are no extraneous layers, or unintended layers with high strain. XRD of (004) planes, taken from orthogonal planes of incidence, is shown in Fig. S5. The most notable characteristics of this structure are the MQW superlattice peaks, the graded buffer, and the lattice-mismatched 0.9 eV GaInAs cell, each labeled in Fig. S5. The superlattice peaks extend outward in both directions of Omega-2Theta from the substrate peak at (0,0). The compressive GaInP graded buffer extends downward from the substrate in Omega-2Theta. A large epilayer tilt, equivalent to the Omega axis of (004) planes, exists in one direction due to the influence of the substrate miscut and dislocation glide through atomically-ordered GaInP graded buffers, discussed elsewhere,[19,20] and is monitored as a metric of dislocation dynamics.[21] Finally, the lattice-mismatched GaInAs subcell is the strong peak at -3100 Omega-2Theta.

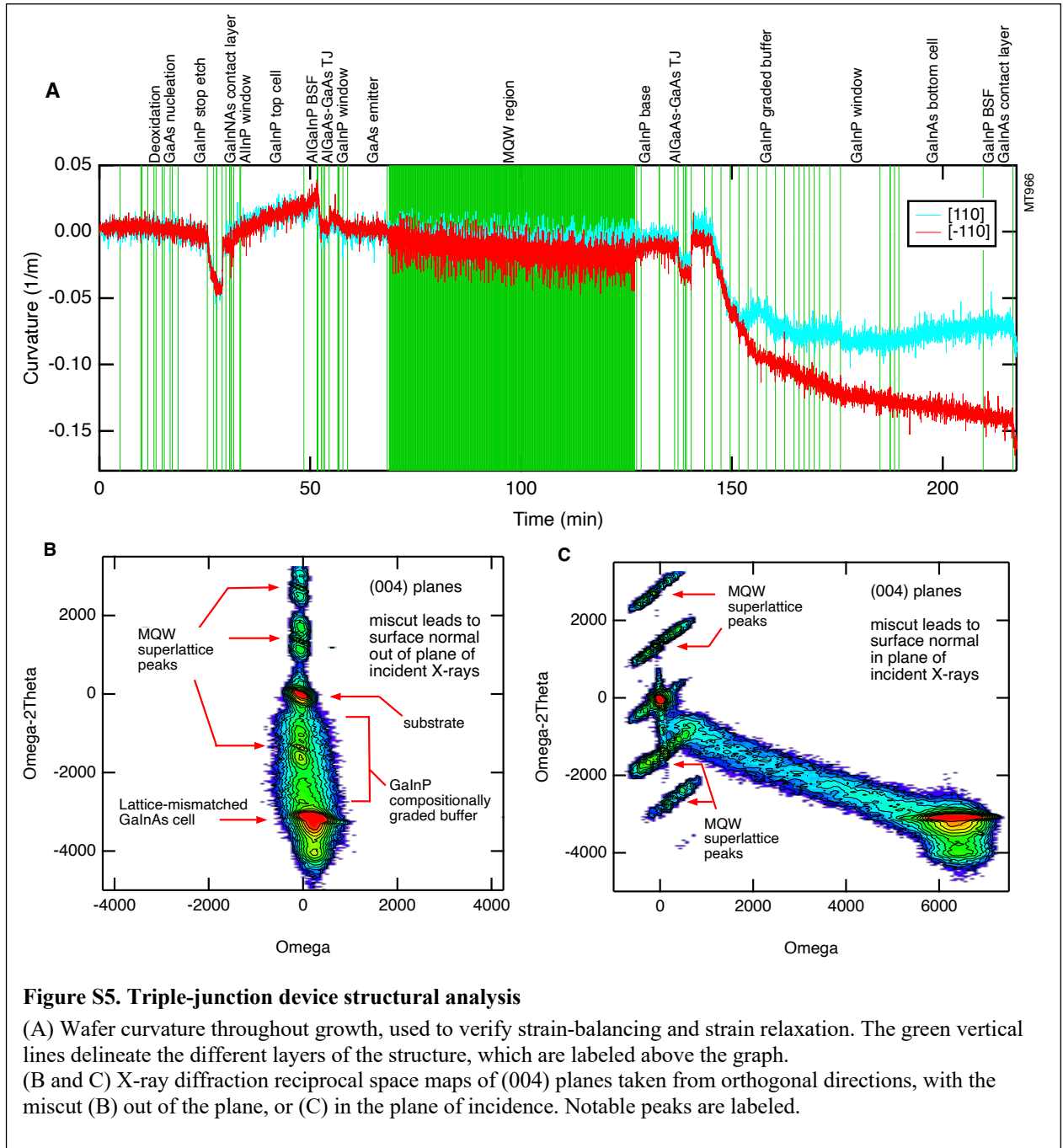

**Figure S5. Triple-junction device structural analysis**
(A) Wafer curvature throughout growth, used to verify strain-balancing and strain relaxation. The green vertical lines delineate the different layers of the structure, which are labeled above the graph.
(B and C) X-ray diffraction reciprocal space maps of (004) planes taken from orthogonal directions, with the miscut (B) out of the plane, or (C) in the plane of incidence. Notable peaks are labeled.

**Triple-junction bandgap modeling results: third subcell**

Figure 1 shows the efficiency of a triple-junction device as a function of the top two cells. The third cell bandgap is optimized for every combination of the top two cells, but is not plotted in Figure 1 to more clearly observe bandgap trends in the top two cells. The third cell bandgap is not constrained in the modeling because as is intended to be a metamorphic cell in a triple-junction inverted metamorphic multijunction design, where the metamorphic bandgap is tunable by altering the compositionally graded buffer.[22,23] The optimized bandgap of the third subcell is plotted in Fig. S6 for both the AM1.5 global and AM0 space spectra.

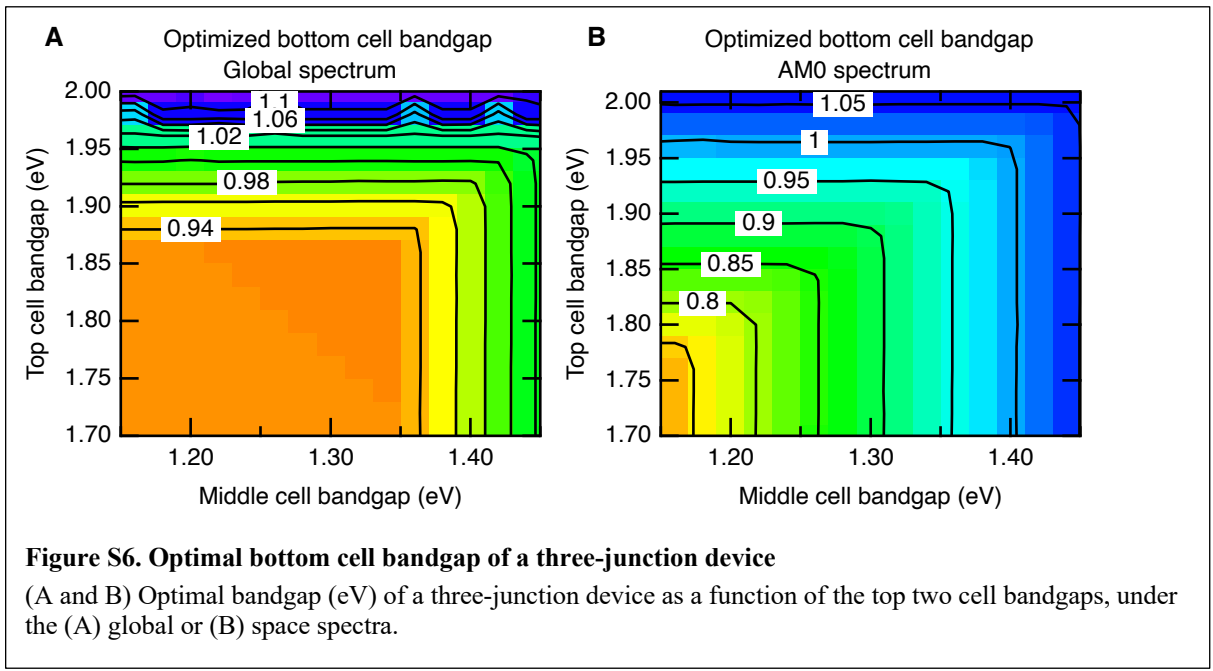

**Figure S6. Optimal bottom cell bandgap of a three-junction device**

(A and B) Optimal bandgap (eV) of a three-junction device as a function of the top two cell bandgaps, under the (A) global or (B) space spectra.

The water absorption gap in the global spectrum leads to a significant difference in optimal bottom junction bandgap compared to the the space spectrum. For many combinations of the top two cells, the optimal bandgap of the third subcell under the global spectrum is about 0.9 eV because the spectral irradiance drops at lower energies due water absorption in the global spectrum. Here the global optimum of the global spectrum is followed similar to the direct spectrum,[24] but a local optimum at 0.7 eV has also been targeted in other CPV 3J work.[25] The space spectrum does not have this water absorption gap, so the third subcell bandgap continually varies with the middle cell bandgap and thus a lower optimal bandgap combination (1.84, 1.26, 0.85 eV) could also be targeted.

# Triple-junction device electroluminescence and external radiative efficiency

Figure 4 shows the subcell dark I-V curves of the triple-junction devices in the paper, highlighting the excellent Woc in each subcell. Equivalently, the external radiative efficiencies (EREs) of each junction are also excellent. Fig. S7 shows the electroluminescence spectra at three different current injections, and the ERE calculated from these spectra. Figure 4 calculates the dark I-V curves from these data using the reciprocity theorem.[26,27] The excellent Woc and ERE at the 1-sun operating current, combined with high collection efficiency at the optimal bandgap combination, leads to the record device performance.

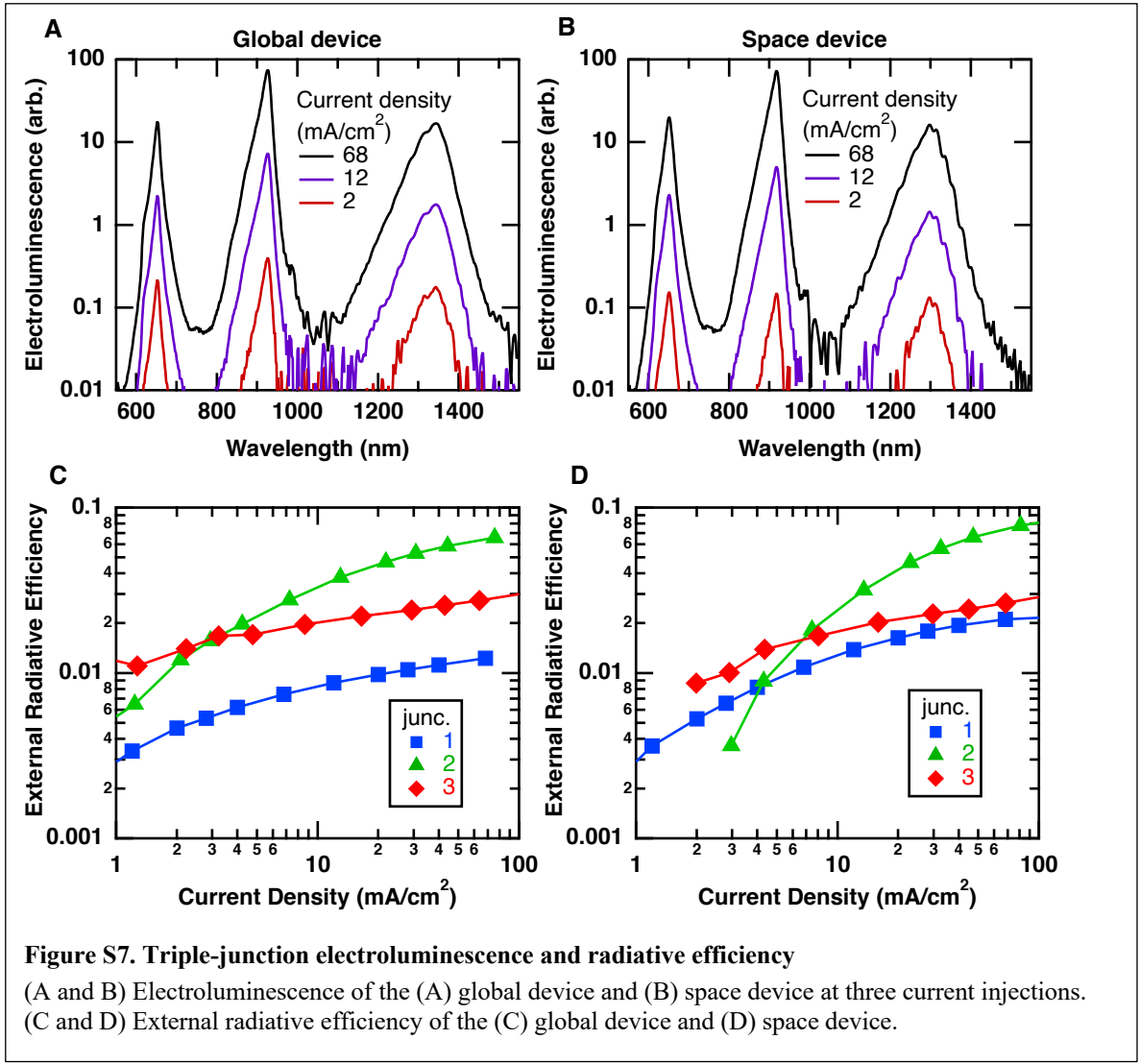

**Figure S7. Triple-junction electroluminescence and radiative efficiency**
(A and B) Electroluminescence of the (A) global device and (B) space device at three current injections.
(C and D) External radiative efficiency of the (C) global device and (D) space device.

# SUPPLEMENTAL REFERENCES